\documentclass[twocolumn]{aastex62}

\def\gch{\gamma_{\rm ch}}

\def\be{\begin{equation}}
\def\ee{\end{equation}}

\usepackage{xcolor}

\begin{document}

\title{Evidence of High-latitude Emission in the Prompt Phase of GRBs: \\
How Far from the Central Engine are the GRBs Produced?} 


\author{Z. Lucas Uhm}
\affiliation{Korea Astronomy and Space Science Institute, Daejeon 34055, Republic of Korea,  \href{mailto:uhm@kasi.re.kr}{uhm@kasi.re.kr}}

\author{Donggeun Tak}
\affiliation{SNU Astronomy Research Center, Seoul National University, Seoul 08826, Republic of Korea}
\affiliation{Deutsches Elektronen-Synchrotron DESY, Platanenallee 6, 15738 Zeuthen, Germany}
\affiliation{Korea Astronomy and Space Science Institute, Daejeon 34055, Republic of Korea}

\author{Bing Zhang}
\affiliation{Department of Physics and Astronomy, University of Nevada, Las Vegas, NV 89154, USA}

\author{Judith Racusin}
\affiliation{Astrophysics Science Division, NASA Goddard Space Flight Center, Greenbelt, MD 20771, USA}

\author{Daniel Kocevski}
\affiliation{Astrophysics Office, ST12, NASA Marshall Space Flight Center, Huntsville, AL 35812, USA}

\author{Sylvain Guiriec}
\affiliation{Department of Physics, The George Washington University, Washington, DC 20052, USA}
\affiliation{Astrophysics Science Division, NASA Goddard Space Flight Center, Greenbelt, MD 20771, USA}

\author{Bin-Bin Zhang}
\affiliation{School of Astronomy and Space Science, Nanjing University, Nanjing 210093, China}

\author{Julie McEnery}
\affiliation{Astrophysics Science Division, NASA Goddard Space Flight Center, Greenbelt, MD 20771, USA}
\affiliation{Department of Physics, University of Maryland, College Park, MD 20742, USA}

\begin{abstract}
One of the difficulties in nailing down the physical mechanism of gamma-ray bursts (GRBs) comes from the fact that there has been no clear observational evidence on how far from the central engine the prompt gamma-rays of GRBs are emitted. Here we present a simple study addressing this question by making use of the ``high-latitude emission'' (HLE). We show that our detailed numerical modeling exhibits a clear signature of HLE in the decaying phase of ``broad pulses'' of GRBs. We show that the HLE can emerge as a prominent spectral break in $F_{\nu}$ spectra and dominate the peak of $\nu F_{\nu}$ spectra even while the ``line-of-sight emission'' (LoSE) is still ongoing. This finding provides a new view of HLE emergence since it has been believed so far that the HLE can show up and dominate the spectra only after the LoSE is turned off. We remark, however, that this ``HLE break'' can be hidden in some broad pulses, depending on the proximity between the peak energies of the LoSE and the HLE. Therefore, this new picture on HLE emergence explains both detection and non-detection of HLE signature in observations of broad pulses. Also, we present three examples of {\em Fermi}-GBM GRBs with broad pulses that exhibit the HLE signature. We show that their gamma-ray emitting region should be located at $\sim 10^{16}$ cm from the central engine, which places a constraint on the GRB models.
\end{abstract}



\section{Introduction}

The gamma-ray bursts (GRBs) are believed to invoke highly relativistic jets with bulk Lorentz factors of a few hundreds \citep{kumarzhang15}. For such a highly relativistic jet, the relativistic beaming and boosting of radiation plays an important role and gives rise to interesting results especially when combined with a spherical geometry of the emitting surface. The photons emitted from a jet location with high latitude, called the ``high-latitude emission'' (HLE), take longer to reach a distant observer and are boosted with a smaller Doppler factor than the photons traveling along the line of sight, called the ``line-of-sight emission'' (LoSE). These two aspects of HLE are known as the ``curvature effect'' of a relativistic spherical jet. It is known that the HLE satisfies a simple relation \citep{kumar00,dermer04,uhm15}, $\hat \alpha = 2 + \hat \beta$, between the temporal index $\hat \alpha$ and the spectral index $\hat \beta$ in the convention of $F_{\nu_{\rm obs}} \propto t_{\rm obs}^{-\hat \alpha}\, \nu_{\rm obs}^{-\hat \beta}$ if the emitter remains a constant Lorentz factor. Here, $F_{\nu_{\rm obs}}$ is the observed spectral energy flux, $t_{\rm obs}$ the observer time, and $\nu_{\rm obs}$ the observed frequency. 
This relation was generalized later for relativistic jets that undergo bulk acceleration ($\hat \alpha > 2 + \hat \beta$) or bulk deceleration ($\hat \alpha < 2 + \hat \beta$) \citep{uhm15}.

The curvature effect of HLE is commonly invoked to account for the steep decay of GRB flares observed in X-rays \citep[e.g.,][]{liang06,uhm16a,jia16} and $\gamma$-rays \citep{Ajello2019}. This effect is also employed to explain the early steep decay during the transition from the prompt emission to the afterglow in the X-ray \citep[e.g.,][]{zhang06, zhangbb09, hascoet12} and GeV energy bands \citep{Ajello2019}. As for the prompt phase of GRBs, several studies \citep{ryde02,kocevski03b,genet09,shenoy13} have investigated the role that the curvature effect has on temporal and spectral properties of individual pulses, but an unambiguous identification of HLE could not be achieved.

The prompt phase of GRBs contains an important observational feature called the ``broad pulses''. Observationally, the broad pulses exhibit two distinct patterns of peak evolution; i.e., the peak-energy ($E_p$) of $\nu F_{\nu}$ spectra shows a ``hard-to-soft'' or a ``flux-tracking'' pattern across the pulses \citep{ford95,norris96,golenetskii83,lu12}. In addition, the light curves of broad pulses in different energy bands exhibit a sequential pattern in their peak time, known as the ``spectral lags'' \citep{norris96,norris00,kocevski03a}; softer emission lags behind harder emission (``positive'' type) in most cases, whereas harder emission can lag behind softer emission (``negative'' type) in some cases. 
The curvature effect of HLE was traditionally suggested as a plausible explanation for the positive type of spectral lags \citep{shen05}, but a detailed study \citep{uhm16b} showed that the HLE cannot give rise to any spectral lags if the spectral shape is softer than $F_{\nu_{\rm obs}} \propto \nu_{\rm obs}^2$. The HLE may produce some spectral lags for a spectral shape harder than this $\nu_{\rm obs}^2$, but the resulting spectral lags are essentially invisible due to the significant flux-level difference between the light curves \citep{uhm16b}.

The complex and intriguing characteristics of broad pulses carry crucial clues to unveil the nature of GRBs. For instance, a series of numerical studies \citep{uhm16b,uhm18} showed that all those features of broad pulses can be successfully reproduced within a single physical picture that invokes a bulk acceleration of the emitting region and that keeps the LoSE ongoing across the production of broad pulses. Also, \cite{lizhang21} found evidence of jet acceleration in an effort of searching for the curvature effect.

Here, we present a simple study that identifies a clear signature of HLE in the decaying phase of broad pulses and provide a new understanding on the HLE emergence. We also present three examples of {\em Fermi}-GBM \citep{meegan09} GRBs that exhibit the HLE signature in their broad pulses.


\section{A simple physical model} \label{section:2}

Following the previous works \citep{uhm16b,uhm18}, 
we adopt a simple physical picture where a thin, relativistic spherical shell expands in space radially. The radiating electrons are distributed uniformly in the shell and emit synchrotron photons \citep{rybicki79} isotropically in the co-moving frame. Then we take fully into account the curvature effect to compute the HLE \citep{uhm15}. We assume a ``Band'' function shape \citep{band93} for the emission spectrum in the co-moving frame since the observed gamma-ray spectra are traditionally fit to this function and since it is a good representation of synchrotron radiation \citep{uhm14,zhangbb16}. The strength of magnetic field $B(r)$ in the emitting region globally decreases as the radius $r$ from the central engine increases, which is expected for a spherical jet traveling in space. We note that this was the essential physical element to explain the low-energy photon index of the Band spectra for the majority of GRBs \citep{uhm14, geng18}. Moreover, the emitting region itself undergoes rapid bulk acceleration \citep{uhm16a,uhm16b} during which the prompt gamma-rays are produced; i.e., the bulk Lorentz factor $\Gamma(r)$ of the region has an increasing profile in radius $r$. Also, the characteristic Lorentz factor $\gch(r)$ of electrons in the co-moving frame is allowed to evolve with radius $r$.

We present three numerical models of broad pulses: Model [u], [v], and [w]. The three models have different $\gch(r)$ profile as described in Appendix~\ref{appendix:1}. Other than $\gch$ profile, we keep all other model parameters the same for the three models, for simplicity. We assume a Band-function shape with typical low- and high-energy photon spectral index $\alpha_{\rm B}=-0.8$ and $\beta_{\rm B}=-2.3$, respectively, for the emission spectrum in the co-moving frame. The number of radiating electrons is assumed to increase at a constant injection rate $R_{\rm inj}=10^{47}$ $\mbox{s}^{-1}$. The bulk Lorentz factor of the jet takes a power-law profile in radius $r$, $\Gamma(r)=\Gamma_0 (r/r_0)^{s}$, with $\Gamma_0=250$, $r_0=10^{15}$ cm, and $s=0.35$, as used in \cite{uhm16b}. We turn on the emission of spherical jet at radius $r_{\rm on}=10^{14}$ cm and turn off its emission at radius $r_{\rm off}=3 \times 10^{16}$ cm. For the given profile of $\Gamma(r)$, this turning-off happens at about $t_{\rm obs}=4.0$ sec. We stress that the LoSE remains ongoing until this turn-off time. The magnetic field strength $B(r)$ in the co-moving frame also takes a power-law profile, $B(r)=B_0 (r/r_0)^{-b}$, with $B_0=30$ G and $b=1.5$ \citep{uhm16b}. We calculate the luminosity distance to GRB for a flat $\Lambda$CDM universe with parameters $\Omega_{\rm m}= 0.31$, $\Omega_{\rm \Lambda}= 0.69$, and $H_0= 68$ km/s/Mpc \citep{Planck16} and take a typical value of redshift $z=1$.


\section{Results of numerical models}

\begin{figure}
\begin{center}
\includegraphics[width=18cm]{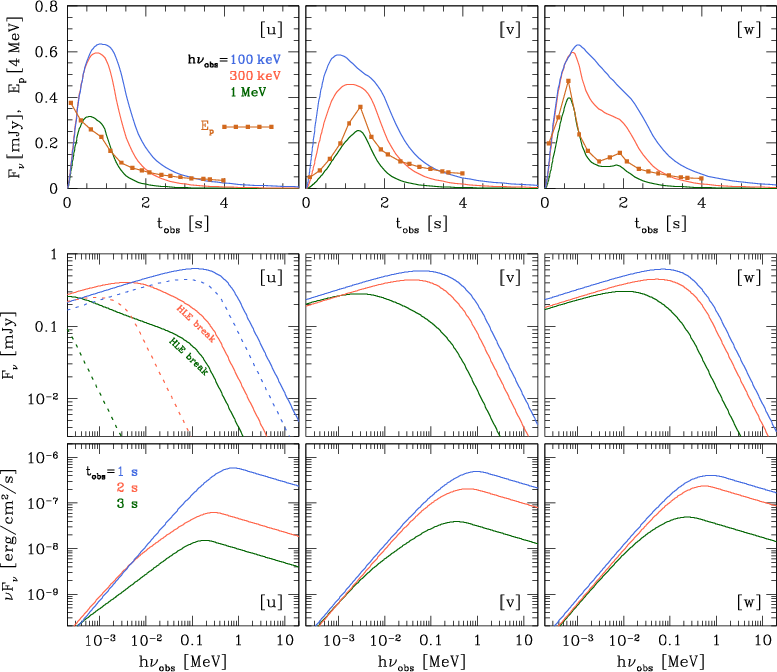}
\caption{
Light curves and time-dependent spectra of our numerical models [u], [v], and [w]. The top panels show the light curves at 100 keV, 300 keV, and 1 MeV, together with temporal evolution of $E_p$ curves. The middle panels show time-dependent spectra at 1 sec, 2 sec, and 3 sec, with the curvature effect of HLE \citep{uhm15} fully included (solid lines) or removed (dotted lines) for model [u]. The $\nu F_{\nu}$ spectra in the bottom panels are derived from the corresponding solid lines in the middle panels. 
}
\label{fig:f1}
\end{center}
\end{figure}

Figure~\ref{fig:f1} shows a modeling result of the three models [u], [v], and [w]. The top panels show the light curves at 100 keV, 300 keV, and 1 MeV, which exhibit both the positive and negative types of spectral lags. The top panels also show the temporal evolution of $E_p$ curves exhibiting both the hard-to-soft and the flux-tracking patterns across the pulses. We plot the $E_p$ points up to $t_{\rm obs}=4.0$ sec during which the LoSE remains ongoing. Note that the breaks in $E_p$ curves are due to the breaks in $\gch$ profiles (see Appendix~\ref{appendix:1}). The lightcurves at 1 MeV also show features linked with the $\gch$ evolution. The middle panels show the time-dependent spectra at 1 sec, 2 sec, and 3 sec (solid lines). In the co-moving frame, we inject a Band-function spectrum with fixed $\alpha_{\rm B}$ and $\beta_{\rm B}$. However, the resulting spectra in the observer frame deviate significantly from this single Band function. Hence, in order to understand this deviation, we repeat the same calculations without considering the curvature effect; the resulting spectra are shown in dotted lines in the middle panel for model [u]. Comparing the solid and dotted lines, one can clearly see that the curvature effect causes the deviation and that the HLE emerges as a prominent additional spectral break in $F_{\nu}$ spectra during the decaying phase of the broad pulses\footnote{We remark that the HLE emergence is modest in model [w] due to a second activity occurring right before 2 sec.}. We stress again that the jet emission is not turned off until about 4 sec in our models and, therefore, the LoSE is still active and dominates the peak of $F_{\nu}$ spectra as it should. The bottom panels show the $\nu F_{\nu}$ spectra directly calculated from the solid lines in the middle panels, in which it is clear that the ``HLE break'' ($\nu_{\rm \scriptscriptstyle HLE}$) in $F_{\nu}$ spectra now becomes the peak energy ($E_p$) in $\nu F_{\nu}$ spectra in the decaying phase of these broad pulses.

We add a remark that an introduction of a break in the profile of $\Gamma(r)$ and/or $B(r)$ induces only a marginal difference and keeps our results unchanged.

If the peak of $\nu F_{\nu}$ spectra is dominated by the HLE in the falling phase of broad pulses, there should exist a simple scaling relation
\begin{equation}
F_{\nu,E_p} \propto E_p^{\,2},
\label{eq:HLE}
\end{equation}
expected from the HLE theory \citep{dermer04,uhm15}. Here, $F_{\nu,E_p}$ is the spectral energy flux $F_{\nu}$ measured at the peak energy $E_p$.

\begin{figure}
\begin{center}
\includegraphics[width=18cm]{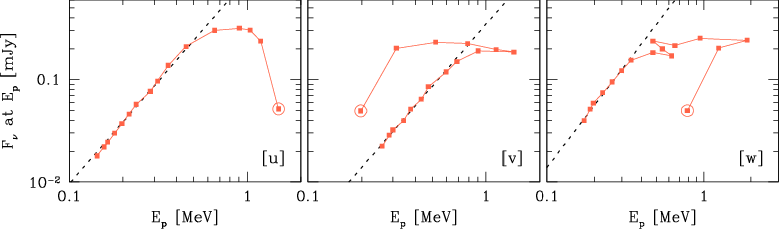}
\caption{
The peak energy $E_p$ vs the spectral flux $F_{\nu}$ at $E_p$ (i.e., $F_{\nu,E_p}$) across the broad pulses of our numerical models [u], [v], and [w]. An open circle in each model marks the first point in the beginning of pulses. The dotted line indicates the relation $F_{\nu,E_p} \propto E_p^{\,2}$ in Equation (\ref{eq:HLE}). 
}
\label{fig:f2}
\end{center}
\end{figure}

In Figure~\ref{fig:f2}, we plot $F_{\nu,E_p}$ against $E_p$ across the broad pulses of three numerical models [u], [v], and [w]. An open circle in each model marks the first point in the beginning of the pulse. One can clearly see that the model curves closely follow Equation (\ref{eq:HLE}) (indicated by the dotted line) in the decaying phase of broad pulses, ascertaining that the peak of $\nu F_{\nu}$ spectra indeed originates from the HLE. This is the clear signature of HLE, produced in our numerical models of broad pulses.


\section{Search of HLE signature in Observations}

{\em Fermi}-GBM \citep{meegan09} has accumulated invaluable observations for the prompt emission of GRBs. In search of the HLE signature above, we analyze a sample of {\em Fermi}-GBM GRBs with relatively clean broad pulses and perform a dedicated time-resolved spectral analysis\footnote{We test three widely-used spectral models with freely varying parameters: the power-law (PL) function, the PL with an exponential cutoff, and the Band function. We determine the best-fit model and its parameters using a proper statistical method.} for each broad pulse; see our companion paper \cite{tak2023} for details.

\begin{figure}
\begin{center}
\includegraphics[width=18cm]{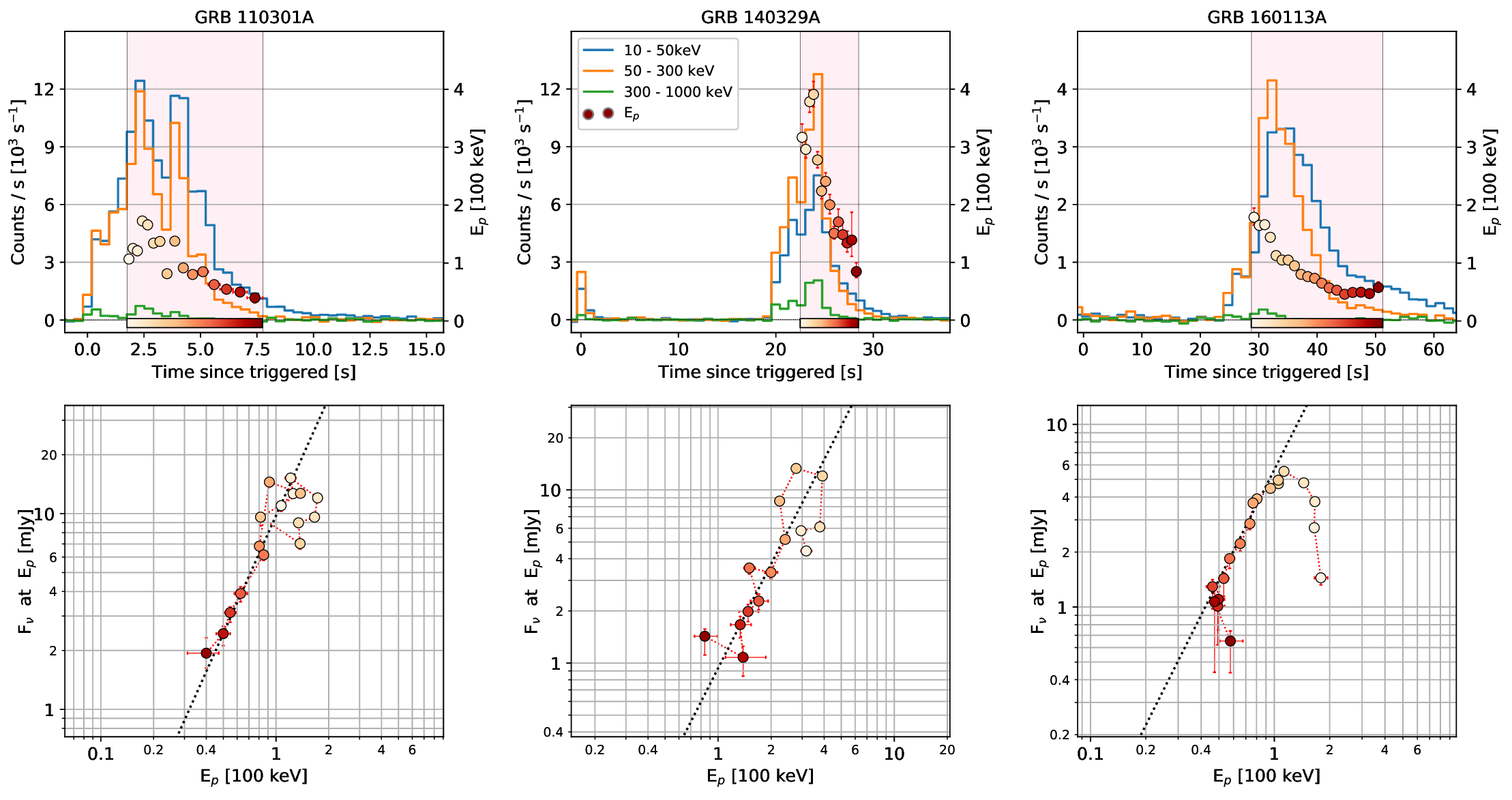}
\caption{
Results of our time-resolved spectral analysis performed on three example broad pulses in GRB 110301A, GRB 140329A, and GRB 160113A. The top panels show the light curves at three different energy bands, together with temporal evolution of $E_p$ points. The bottom panels show the $E_p$ vs $F_{\nu,E_p}$ points obtained from the analysis. The dotted line indicates the relation $F_{\nu,E_p} \propto E_p^{\,2}$ in Equation (\ref{eq:HLE}). The color gradient in the bottom panels is in accordance with the color gradient encoded in $E_p$ points in the top panels.
}
\label{fig:f3}
\end{center}
\end{figure}

In Figure~\ref{fig:f3}, we present three examples: GRB 110301A, GRB 140329A, and GRB 160113A. The top panels show the light curves at three different energy bands, together with the temporal evolution of $E_p$ curves. The bottom panels show $E_p$ vs $F_{\nu,E_p}$ obtained from the time-resolved spectral analysis. The dotted line indicates the theoretical HLE relation in Equation (\ref{eq:HLE}). As one can see, the $F_{\nu,E_p}$ - $E_p$ points obtained from the time-resolved analysis of the three bursts are in good agreement with Equation (\ref{eq:HLE}) in the decaying phase of their broad pulse, implying that the HLE signature is indeed identified. The color gradient used in the bottom panels is in accordance with the color gradient encoded in $E_p$ points in the top panels, which helps locate where in the pulse the HLE signature starts to show up.

While numerous studies have explored the correlation between $E_p$ and flux, this HLE signature we investigate has not been reported. This is primarily because most empirical relations are derived during the brightest phase of GRBs, whereas our study focuses on the falling phase of the broad pulses. Further details on the outcomes can be found in \cite{tak2023}.


\section{Conclusions and Discussion}

In this paper, we showed that the HLE can imprint a clear spectral signature in prompt-emission gamma-ray spectra (as an additional spectral break $\nu_{\rm \scriptscriptstyle HLE}$ in $F_{\nu}$ spectra and as the peak energy in $\nu F_{\nu}$ spectra) even in the existence of ongoing LoSE. This result provides a new view regarding the HLE, because it has been believed so far that the HLE can show up and dominate the spectra only after the LoSE is turned off.

We remark, however, that the HLE spectral break is not required to appear in all broad pulses. It is because the HLE break can be buried under the ongoing LoSE component when the peak energy of LoSE (at a given time) is not far below that of HLE (emitted at earlier times but belonging to the same equal-arrival-time surface). Therefore, this new perspective on the HLE emergence provides flexibility to explain both the detection and the non-detection of HLE signature in observations of broad pulses.

The location of the LoSE peak depends on the physical parameters in the emitting region, and in our models, it is roughly given by
\begin{equation}
    \nu_{\rm \scriptscriptstyle LoSE} \propto \Gamma B \gch^2 \propto r^{s-b+2g} \propto t_{\rm obs}^{(s-b+2g)/(1-2s)},
\end{equation}
where we assumed $\gch(r) \propto r^g$ and used the relation $t_{\rm obs} \propto r/\Gamma^2 \propto r^{1-2s}$. On the other hand, the HLE break is expected to evolve in time as
\begin{equation}
    \nu_{\rm \scriptscriptstyle HLE} \propto t_{\rm obs}^{-1},
\end{equation}
obeying the property of HLE. Therefore, we have the ratio of the two frequencies as
\begin{equation}
    \frac{\nu_{\rm \scriptscriptstyle LoSE}}{\nu_{\rm \scriptscriptstyle HLE}} \propto t_{\rm obs}^{-(s+b-2g-1)/(1-2s)}.
\end{equation}
When $\nu_{\rm \scriptscriptstyle LoSE}$ is sufficiently smaller than $\nu_{\rm \scriptscriptstyle HLE}$ (i.e., $\nu_{\rm \scriptscriptstyle LoSE}/\nu_{\rm \scriptscriptstyle HLE} \ll 1$), one can expect that the HLE break shows up in $F_\nu$ spectra and the HLE dominates the peak of $\nu F_\nu$ spectra.

The falling phase of the lightcurves in our numerical models shows a steep decay and satisfies the well-known closure relation, $\hat{\alpha} = 2+\hat{\beta}$; specifically, $\hat{\alpha} \sim 3.3$ and $\hat{\beta} \sim 1.3$. This suggests that, even when the LoSE remains ongoing, the early steep decay phase following the prompt emission can be interpreted as the HLE of the prompt emission tail.

In this paper, we also presented three {\em Fermi}-GBM broad pulses that exhibit the HLE scaling relation between $E_p$ and $F_{\nu,E_p}$ (Equation~\ref{eq:HLE}) in their decaying phase.

The HLE signature observed in some broad pulses leads to important implication regarding the emission radius of GRBs. The HLE emitted at radius $r$ is received at an observer time given roughly by $t_{\rm obs} \sim r/(2c\,\Gamma^2)$ like in the case of LoSE, which yields 
\begin{equation}
r \sim 2c\, \Gamma^2\, t_{\rm obs} =
(1.6\times 10^{16}\, \mbox{cm}) \left(\frac{\Gamma}{300} \right)^2\, \left(\frac{t_{\rm obs}}{3\, \mbox{s}} \right),
\end{equation}
where $c$ is the speed of light. The duration of broad pulses in our examples is tens of seconds, and therefore the gamma-ray emitting region of those GRBs with HLE signature should be located at $\sim 10^{16}$ cm from the central engine for a typical value of $\Gamma=300$ \citep{Ghirlanda2018}. For a broad range of $\Gamma$ spanning from 100 to 1000 \citep[e.g.,][]{piran99}, the emission radius varies from $\sim 10^{15}$ cm to $\sim 10^{17}$ cm.

This inference of the emission radius is independent of details of our modeling and sheds light on constraining the GRB models. The estimated large emission radius is consistent with the ICMART model \citep{zhangyan11}, which invokes collision-induced magnetic dissipation as the origin of GRB prompt emission. In addition, this implies that some GRB models, such as classical photospheric emission models and small-radii internal shock models, face challenges\footnote{The HLE exists in these models as well but the related timescales are much shorter than the duration of broad pulses by orders of magnitude. Therefore, the scaling relation in Equation (\ref{eq:HLE}) cannot be observed in the decaying phase of broad pulses unless their central engine behaves in a specific manner to produce this scaling relation, which is too contrived.} and demand modifications to account for those GRBs exhibiting the HLE signature.

In short, we identified a clear signature of HLE in the prompt phase of GRBs both theoretically and observationally. Also, we presented a unique constraint on the validity of the competing GRB models.

\acknowledgments

This work was supported by the National Research Foundation of Korea (NRF) grant, No. 2021M3F7A1084525, funded by the Korea government (MSIT). B.B.Z acknowledges the support by the National Key Research and Development Programs of China (2018YFA0404204, 2022YFF0711404, 2022SKA0130102), the National Natural Science Foundation of China (Grant Nos. 11833003, U2038105, 12121003), the science research grants from the China Manned Space Project with NO.CMS-CSST-2021-B11, and the Program for Innovative Talents, Entrepreneur in Jiangsu.

%
%


\appendix

\section{Profile of characteristic Lorentz factor $\gch$ of electrons} \label{appendix:1}

\begin{figure}
\begin{center}
\includegraphics[width=10cm]{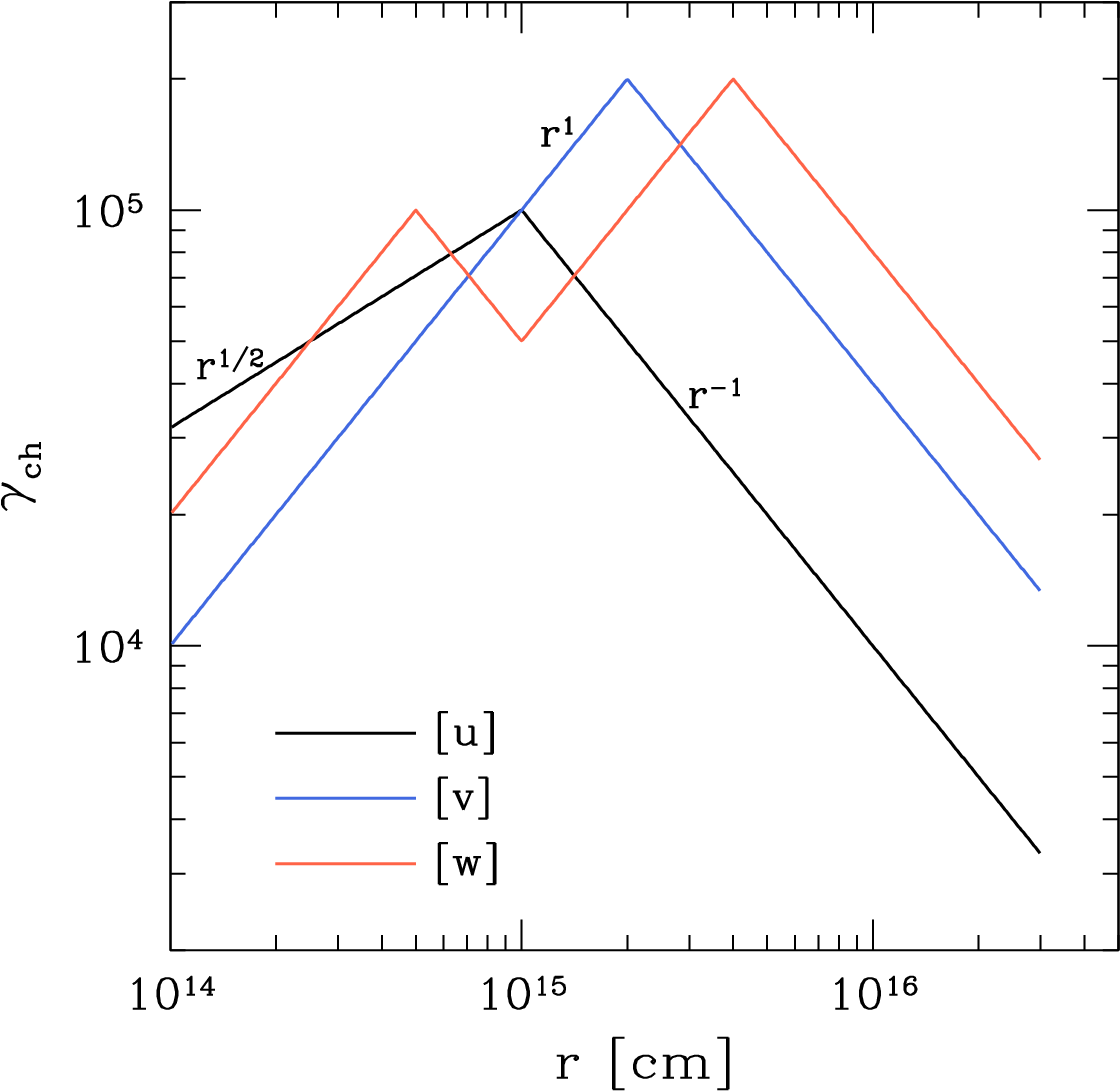}
\caption{
Profile of characteristic Lorentz factor $\gamma_{\rm ch}$ of electrons in our numerical models [u], [v], and [w].
}
\label{fig:fS1}
\end{center}
\end{figure}

The characteristic Lorentz factor $\gch$ of electrons evolves in radius $r$ in our models. We expect that the rate of dissipation of internal energy will determine the values of $\gch$. If we think of an episode of energy dissipation, the dissipation may initially occur in an explosive manner such that $\gch$ increases initially, which could be followed by a fading phase of dissipation with $\gch$ decreasing. In this picture, it is plausible to assume that the $\gch$-profile evolves in radius $r$ as the emitting shell expands in space. The model [u] has a broken power-law profile
\begin{eqnarray}
\gch(r) =
\left\{
\begin{array}{ll}
\gch^0 (r/r_0)^{g_1}   \quad & \mbox{if} \quad r < r_0, \\
\gch^0 (r/r_0)^{g_2}  \quad & \mbox{if} \quad r \geq r_0,
\end{array}
\right.
\end{eqnarray}
with $\gch^0 = 10^5$, $r_0 = 10^{15}$ cm, $g_1 = 1/2$, and $g_2=-1$. The model [v] also takes the same form of broken power-law but with $\gch^0 =2\times 10^5$, $r_0 =2\times 10^{15}$ cm, $g_1= 1$, and $g_2 = -1$. The model [w] has a profile made of four power-law segments
\begin{eqnarray}
\gch(r) =
\left\{
\begin{array}{ll}
\gch^0 (r/r_1)        \quad & \mbox{if} \quad          r < r_1, \\
\gch^0 (r/r_1)^{-1}   \quad & \mbox{if} \quad r_1 \leq r < r_2, \\
2\gch^0 (r/r_3)       \quad & \mbox{if} \quad r_2 \leq r < r_3, \\
2\gch^0 (r/r_3)^{-1}  \quad & \mbox{if} \quad r_3 \leq r,
\end{array}
\right.
\end{eqnarray}
with $\gch^0 = 10^5$, $r_1 =5\times 10^{14}$ cm, $r_2 =10^{15}$ cm, and $r_3 =4\times 10^{15}$ cm. 
These three $\gch$ profiles are shown in Figure~\ref{fig:fS1}.

%
%


%
%

\clearpage

\end{document}